# Super-resolution image projection over an extended depth of field using a diffractive decoder

Hanlong Chen[1,2,3†], Çağatay Işıl[1,2,3†], Tianyi Gan[1], Mona Jarrahi[1], Aydogan Ozcan[1,2,3*]

1 Electrical and Computer Engineering Department, University of California, Los Angeles, California 90095, USA

2 Bioengineering Department, University of California, Los Angeles, California 90095, USA

3 California NanoSystems Institute (CNSI), University of California, Los Angeles, California 90095, USA

[†]These authors contributed equally to this work.

[*]Corresponding author: ozcan@ucla.edu

## Abstract

Image projection systems must be efficient in data storage, computation and transmission while maintaining a large space-bandwidth-product (SBP) at their output. Here, we introduce a hybrid image projection system that achieves extended depth-of-field (DOF) with improved resolution, combining a convolutional neural network (CNN)-based digital encoder with an all-optical diffractive decoder. A CNN-based encoder compresses input images into compact phase representations, which are subsequently displayed by a low-resolution (LR) projector and processed by an analog diffractive decoder for all-optical image reconstruction. This optical decoder is completely passive, designed to synthesize pixel super-resolved image projections that feature an extended DOF while eliminating the need for additional power consumption for super-resolved image reconstruction. Our pixel super-resolution (PSR) image projection system demonstrates high-fidelity image synthesis over an extended DOF of ~$267\lambda$, where $\lambda$ is the illumination wavelength, concurrently offering up to ~16-fold SBP improvement at each lateral plane. The proof of concept of this approach is validated through an experiment conducted in the THz spectrum, and the system is scalable across different parts of the electromagnetic spectrum. This image projection architecture can reduce data storage and transmission requirements for display systems without imposing additional power constraints on the optical decoder. Beyond extended DOF PSR image projection, the underlying principles of this approach can be extended to various applications, including optical metrology and microscopy.

## Introduction

Augmented and virtual reality (AR/VR) systems hold immense promise across diverse fields, including education, entertainment, and healthcare[1,2]. These systems aim to deliver immersive user experiences. Near-eye displays, while crucial for achieving this objective, have presented a hurdle in realizing the full potential of AR/VR technology. A critical issue plaguing most near-eye displays, including stereoscopic, autostereoscopic, and multi-view systems, is the well-documented discomfort and fatigue experienced by users[3,4]. This primarily stems from the mismatch between vergence and accommodation cues of the eye. Holographic displays, in contrast to these widely adopted approaches, possess the ability to offer all depth cues of the human visual system. This capability can potentially alleviate eye strain and visual discomfort associated with vergence-accommodation conflict[4-9], which is important for AR/VR systems. However, the space-bandwidth product (SBP) of holographic displays is constrained by the limited pixel count of spatial light modulators (SLMs). In addition to this, further research into advanced computer-generated holography (CGH) techniques and supporting hardware is crucial for enabling real-time, full-parallax holographic displays with low power consumption[7,8,10,11]. Advances in machine learning techniques have also been applied to improve hologram calculation in terms of accuracy and speed[12-21], as well as to enhance the SBP of SLMs[22,23] through passive surfaces. An alternative strategy involves computing holograms on a remote server and transmitting them to local receivers with limited, low-cost processors[4]. However, hologram compression presents challenges, as conventional image compression techniques like JPEG (Joint Photographic Experts Group) inevitably sacrifice the fine details critical for synthesizing high-quality holographic images. These limitations highlight the need for innovative image projection approaches that can simultaneously address the current challenges associated with data transmission, processing power, and restricted SBP in display technologies.

Here, we present a deep learning-enabled image projection approach that synergistically integrates a digital encoder with a passive all-optical decoder, achieving both extended depth of field (DOF) and pixel super-resolution (PSR). This hybrid architecture employs a convolutional neural network (CNN) to efficiently encode high-resolution image information into compact phase-only representations, which are subsequently displayed by a low-resolution (LR) projector and processed by a passive diffractive decoder to synthesize super-resolved images over an extended DOF. We demonstrate the capabilities of our system through comprehensive numerical simulations and proof-of-concept experiments using terahertz radiation, showcasing its ability to project images over a DOF of $\sim 267\lambda$ with approximately 16-fold SBP improvement compared to the SBP of the input projector, where $\lambda$ is the illumination wavelength. This digital encoder – analog decoder architecture might help us address data storage and transmission requirements while reducing computational costs since it incurs no additional power consumption or latency at the user end, leveraging the intrinsic advantages of its all-optical passive decoder. This hybrid super-resolution image projection system with its extended DOF might find various applications, including in optical metrology and microscopy.

## Results

The hybrid architecture of our extended DOF PSR image projection system integrates a digital encoder with an all-optical passive diffractive decoder, as depicted in **Figure 1**. The electronic encoder, shown in **Figure 1a**, employs a lightweight CNN-based encoder to process the input image through a sequence of convolutional layers (**Figure 1a**). Each layer is equipped with instance normalization[24] and Parametric Rectified Linear Unit (PReLU) activation[25]. In our proof of concept demonstration, this CNN-based encoder is designed to compress grayscale images with $24 \times 24$ pixels into compact phase representations with $\frac{24}{k} \times \frac{24}{k}$ pixels where $k = 6$ in **Figure 1a** – i.e., a reduction of 36-fold in the total number of pixels per image. The all-optical analog decoder is composed of three diffractive layers ($L = 3$), illustrated in **Figure 1b**, and it is jointly designed/optimized to process the encoded compact phase patterns to synthesize high-resolution images over an extended DOF (EDOF), denoted by $z_e$. Stated differently, the EDOF PSR projection framework performs a continuous volumetric projection while improving the effective SBP of the reconstructed images, which is illustrated using 2D projection slices at the selected discrete positions represented by $z_{3,1}, z_{3,2}, ..., z_{3,i}, ..., z_{3,n}$, with the corresponding cross-sectional images denoted as $I_1, I_2, ..., I_i, ..., I_n$, respectively. To train our hybrid models, we utilized the EMNIST[26] handwritten letter dataset, comprising 80,000 training, 8,800 validation, and 14,800 testing images, each with $24 \times 24$ pixels, and employed the AdamW optimizer, as detailed in the **Methods** section.

**Figure 2** compares the image projection performances of diffractive projection systems with different numbers of passive diffractive layers (L = 1, 3) trained for different PSR factors (k = 4, 6, 8). Each configuration, represented by a distinct row in **Figure 2**, is separately trained (see **Supplementary Figure 1** for the decoders' phase profiles corresponding to each design), and each design is blindly evaluated using the test dataset not seen by the models during the training phase. As an example, a test image of handwritten "*P*" is processed through the diffractive image projection systems and the resulting continuous volumetric projections are presented using lateral cross-sections at selected propagation distances, which showcase the system's ability to maintain high-fidelity projections throughout an extended DOF, axially spanning $z_e = \sim 266.85\lambda$. For instance, the three-layer-based decoder configurations (L = 3) effectively maintain a similar image quality across the entire axial depth range at PSR factors of k = 4, 6, and 8. On the other hand, single-layer decoder configurations ($L = 1$) show a more pronounced trade-off between resolution and DOF performance (see **Fig. 2**). Notably, configurations with a free-space decoder (L = 0 – i.e., ***no*** diffractive decoder) present significantly deteriorated images synthesized across the same DOF, confirming the essential role of the diffractive decoder layers jointly optimized with the digital encoder at the front-end.

We further explored our hybrid projection system by demonstrating its *external generalization* capabilities in **Fig. 3**. Horizontal and vertical gratings, which differ from the training and test datasets used in the previous results, were employed to test the external generalization of our extended DOF PSR image projection system. As illustrated in **Fig. 3**, despite being training using solely the EMNIST dataset, the three-layer-based decoder models accurately synthesized images of gratings with a linewidth of $10.6\lambda$ (~7.94 mm) over an extended DOF $\{z_3 \in \mathbb{R} \mid 150 \leq z_3 \leq 350 \}$ for different PSR factors of k = 4, 6, 8. Notably, the single-layer designs for $k = 4$ and 6 also performed image projections over an extended DOF with some minor degradation in image quality compared to three-layer decoder results. **Figure 3** further reveals that our extended DOF method achieves up to ~16-fold SBP improvement compared to the input/encoding plane, in addition to proving external generalization of the presented framework. Specifically, the model trained for $k = 8$ successfully projected gratings with a linewidth of $10.6\lambda$ (~7.94 mm), translating to an effective pixel size of $5.3\lambda$ (~3.97 mm) at the output plane. These results are obtained using an effective pixel size of 16 mm (~$21.35\lambda$) at the encoded phase plane, corresponding to a ~16-fold improvement in the effective SBP[22,27]. Even smaller linewidths can be imaged through this diffractive platform by specifically training it, end-to-end, for the projection of densely-packed fringe patterns as illustrated in **Supplementary Figure 2.**

**Figure 4** further presents quantitative and visual analysis of our hybrid image projection approach across varying projection distances and depths. For this purpose, different three-layer designs ($L = 3$) targeting a PSR factor of $k = 6$ were trained for various $z_3^{tr}$ ranges and tested using different $z_3^{te}$ ranges. **Figure 4a** elucidates the impact of expanding the depth range $z_e$ to 50 mm (~$66.71\lambda$), 100 mm (~$133.43\lambda$), 150 mm (~$200.14\lambda$), and 200 mm (~$266.85\lambda$), while maintaining a fixed starting point $z_{3,1} = 150$ mm (~$200.14\lambda$). As the DOF broadens, the peak average structural similarity index measure (SSIM) gradually declines, indicating a natural trade-off between the DOF and image projection fidelity. In contrast, an image projection system with a free-space-based decoder (i.e., $L = 0$, trained and evaluated over an equivalent axial range), shown with the dashed purple curve in **Figure 4a**, exhibits significantly degraded performance, once again highlighting the crucial role of the passive diffractive decoder in extending the DOF while maintaining high image projection fidelity. Similarly, **Figs. 4b-c** illustrate the influence of depth range placement and its axial width on the image projection performance. While narrower DOF ranges yield higher SSIM values at the output projection volume, the system, in general, maintains high image quality and discernible features across all the tested DOF ranges reported in **Fig. 4**.

In addition to these numerical analyses, our EDOF PSR image projection approach was also validated in a proof-of-concept experiment. We designed a model with a single-layer diffractive decoder trained for $z_3^{tr} = 135 - 185$ mm ($180.12\lambda - 246.84\lambda$), corresponding to a $z_e = 50\ mm$ (~$66.71\lambda$) and a PSR factor of $k = 6$. To have a misalignment resilient and efficient design, an additional power efficiency-related loss term was used and 3D random displacements were applied to the diffractive decoder surface during the training phase; see the **Methods** section for details. As depicted in **Fig. 5**, we used a THz detector and a continuous-wave THz illumination source ($\lambda =$ ~0.75 mm) in our experiments after the 3D fabrication of the single-layer diffractive surface and the LR encoded-phase profiles, all of which were fabricated using a 3D printer (see the **Methods** section). **Figure 6** illustrates the experimental and the corresponding numerical results for different positions ($z_3^{te}$) of the output plane, including 135 mm (~$180.12\lambda$), 160 mm (~$213.48\lambda$), and 185 mm (~$246.84\lambda$). These measurements support the accuracy and resilience of our image projection method, providing enhanced resolution over an extended depth. These experimental results are particularly important given the simplicity of the single-layer diffractive decoder, highlighting the system's efficiency in terms of both performance and resource utilization.

## Discussion

We presented a deep learning-enabled hybrid image projection system with enhanced DOF and resolution. This diffractive image projection system, integrating a CNN-based digital encoder and an all-optical passive decoder, reduces computational costs across various projection distances for a single wavefront modulator while also enhancing both the DOF and the spatial resolution. The all-optical decoder part of the presented approach can be fabricated using passive optical materials. Moreover, this framework enhances data storage and transmission efficiency by encoding image information into compact phase patterns with significantly compressed dimensions. These patterns are subsequently processed by an all-optical decoder, resulting in the projection of high-resolution images with extended DOF.

To contextualize our approach, classical optical systems typically extend their DOF by reducing the numerical aperture (NA), which imposes a fundamental trade-off between axial depth, lateral resolution, and optical throughput. Our hybrid system circumvents this compromise, using a learned diffractive decoder to process information across the full aperture, thereby achieving both pixel super-resolution and an extended DOF, while maintaining high throughput. Furthermore, this diffractive approach offers a scalable solution; all-optical decoders can be fabricated over large areas using methods like 3D printing, two-photon polymerization and wafer-scale optical lithography.

Further advances in this technology promise to unlock new paradigms across diverse fields ranging from display technologies to advanced scientific instrumentation and optical metrology. For example, deep learning-based metrology methods have recently attracted considerable attention[28], owing to their advantages over traditional model-based approaches. Our diffractive image projection system, combining a CNN-based encoder with an all-optical decoder, might potentially offer new types of solutions to optical metrology tasks. To exemplify this opportunity, we trained an extended DOF PSR image projection system specifically on the projection of fringe patterns with varying linewidths and directions. As illustrated in **Supplementary Fig. 2**, our system successfully projects the desired fringe patterns with varying linewidths across an extended DOF. The presented fringe projection framework, leveraging its enhanced DOF capabilities, can provide an improved approach for non-contact/remote 3D surface measurements, potentially surpassing the performance of conventional metrology methods.

The presented extended DOF PSR image projection approach can also operate across different parts of the electromagnetic spectrum. The designed models can be scaled to any part of the spectrum, including the visible and IR wavelengths, by physically scaling the dimensions of the spatial features with respect to the wavelength of light. To meet the challenge of such smaller-scale features at shorter wavelengths, the all-optical decoder component of the designed model can be fabricated using nano-fabrication methods such as wafer-scale optical lithography and two-photon polymerization[29–32]. Furthermore, precise alignment of the diffractive layers of the decoder can pose additional challenges for operation at shorter wavelengths. Some of these challenges can be mitigated by optimizing misalignment-resilient designs, implemented through a numerical "vaccination" design strategy[33,34], by introducing potential errors and variations in, e.g., size/shape/orientation of the diffractive layers during the training process. The same approach can also be applied to reduce sensitivity to potential fabrication errors in the phase modulation features of the resulting diffractive layers. This adaptability for operation at different parts of the electromagnetic spectrum could facilitate seamless integration with existing optical setups, potentially enhancing their capabilities without necessitating significant modifications. For instance, in the context of structured illumination microscopy (SIM), our system's ability to precisely generate dynamic complex patterns over an extended DOF (e.g., **Supplementary Fig. 2**) could enhance resolution and contrast in various imaging applications, potentially enabling high-throughput 3D structures and dynamic processes to be imaged.

In summary, our EDOF PSR image projection system, enabled by the symbiosis of deep learning with diffractive optical processors, represents a significant advance toward versatile and high-performance light shaping, with far-reaching implications for fields that demand precise lateral and depth control of light, including microscopy and metrology.

## Methods

### Digital encoder design

Our digital encoder, depicted in **Figure 1a**, consists of a series of convolutional layers followed by a fully connected layer. It accepts a $24 \times 24$ grayscale image as input. Each convolutional layer applies its filters to the input, followed by normalization and a PReLU[35] activation. Max-pooling operations between the layers progressively reduce the spatial dimensions while increasing the feature depth. As the input progresses through the network, the spatial dimensions are reduced while the number of channels increases. The final convolutional layer yields a $3 \times 3 \times 128$ feature map, which is then flattened and processed by a fully connected layer with sigmoid activation, producing a compact encoded representation ($\frac{24}{k} \times \frac{24}{k}$ pixels, where a PSR factor of $k = 6$ is used in **Figure 1a**). This encoding operation enables the digital encoder to efficiently compress the essential information of the input image into a phase-only format optimized for subsequent optical decoding.

### Optical decoder design

Our diffractive decoder employs passive diffractive layers, $L \geq 1$, and each layer $l$ is then subdivided into a grid of diffractive features[22,36–38]. These diffractive features, characterized by their transmittance coefficients $t_l[m, n]$, serve as the fundamental building blocks of the all-optical decoder. The transmittance of an individual diffractive feature is defined as follows:

$$t_l[m, n] = \exp\left(j \frac{2\pi}{\lambda} (\tau(\lambda) - n_a) h_l[m, n]\right) \qquad (1)$$

In this expression, the wavelength-dependent complex refractive index $\tau(\lambda) = n(\lambda) + j\kappa(\lambda)$ and the refractive index of the surrounding medium $n_a$ determine the phase modulation imparted by each diffractive feature. A key parameter in the modulation process is the physical thickness of each diffractive feature, i.e., $h_l[m,n]$, which is computed using an auxiliary variable $o_l[m,n]$:

$$h_l[m,n] = \frac{\tanh(o_l[m,n]) + 1}{2}(h_m - h_b) + h_b \tag{2}$$

This formulation allows for a continuous spectrum of thickness values, bounded by $h_b$ and $h_m$, which can be tuned during the optimization process to achieve desired optical properties. These diffractive features give rise to a transmission modulation function $T_l(x, y)$, described by:

$$T_l(x,y) = \sum_m \sum_n t_l[m,n] p_l(x - mW_x, y - nW_y) \tag{3}$$

where the 2D rectangular sampling kernel $p_l(x, y)$ is defined as:

$$p_l(x,y) = \begin{cases} 1, & |x| < \frac{W_x}{2}, |y| < \frac{W_y}{2} \\ 0, & otherwise \end{cases} \tag{4}$$

To model the propagation of spatially and temporally coherent light propagating through this system, the framework employs the angular spectrum method[37], which is based on the fast Fourier transform-based implementation of the Rayleigh-Sommerfeld diffraction integral. This propagation is mathematically expressed as:

$$U_{l+1}(x,y) = U'_l(x,y) * w(x, y, z_{l+1} - z_l) \tag{5}$$

where $U'_l(x, y) = U_l(x, y) T_l(x, y)$ is the output field of layer $l$, and the term $z_{l+1} - z_l$ represents the axial propagation distance between the diffractive layer $l$ and the next diffractive layer $l + 1$. The propagation kernel $w(x, y, z)$ is given by:

$$w(x,y,z) = \frac{z}{r^2}\left(\frac{1}{2\pi r} + \frac{1}{j\lambda}\right) \exp\left(j\frac{2\pi r}{\lambda}\right) \tag{6}$$

with $r = \sqrt{x^2 + y^2 + z^2}$.

**Datasets and image preprocessing**

The EMNIST handwritten letters dataset, consisting of 88,800 training and 14,800 testing samples, was used to train our numerical designs. We allocated 8,800 samples from the training set for validation, yielding 80,000 training, 8,800 validation, and 14,800 testing samples. Handwritten letter images ($28 \times 28$ pixels) from this dataset were resized to $24 \times 24$ pixels using the Bicubic downsampling method. For the experimental phase, we utilized the same dataset to train the hybrid image projection system. To evaluate its performance, we tested the model using handwritten letter images that are different from those in the training dataset.

**Optical decoder parameters**

Our design incorporated diffractive layers spanning $133.42\lambda \times 133.42\lambda$, each discretized into a $200 \times 200$ element grid. The size of diffractive features and the sampling period of the propagation model were set to $0.667\lambda$. The input and output fields of views of the decoder are $96 \times 96$ pixels ($64.04\lambda \times 64.04\lambda$), and aliasing was prevented through zero padding to 400 × 400 pixels. The transmittance coefficient of each diffractive element was expressed as:

$$t_l[m,n] = \exp(j\theta_l[m,n]) = \exp\left(j\frac{2\pi}{\lambda}(n(\lambda)-n_a)h_l[m,n]\right) \quad (7)$$

Here, the phase coefficients $\theta_l[m,n]$ were initialized to zero and subsequently refined through stochastic gradient-based error backpropagation[39].

We explored different decoder configurations with $L = 1, 3$ and $5$ transmissive layers; for comparison, we also trained $L = 0$ designs without a diffractive decoder layer. By utilizing a grid search method, we determined the optimal distance from the input plane to the first diffractive layer ($z_1$), the inter-layer spacing between the diffractive layers ($z_2$), and the distance from the final layer to the first output plane ($z_{3,1}$), as illustrated in **Fig. 1b** and **Table 1**. The final configuration was determined by maximizing the average peak signal-to-noise ratio (PSNR) across the validation dataset, ensuring optimized and robust performance under varied input conditions.

In our experimental results, a monochromatic terahertz illumination source emitting at $\lambda \approx 0.75\ mm$ was used. Our experiments focused on a single-layer ($L = 1$) diffractive decoder, with specific axial distances also detailed in Table 1.

| Distance | $z_1$ | $z_2$ | $z_{3,1}$ | $z_e$ |
|---|---|---|---|---|
| Numerical | $10\ mm\ (13\lambda)$ | $50\ mm\ (66\lambda)$ | $150\ mm\ (200\lambda)$ | $200\ mm\ (267\lambda)$ |
| Experimental | $10\ mm\ (13\lambda)$ | N/A | $155\ mm\ (207\lambda)$ | $50\ mm\ (66\lambda)$ |

**Table 1** The axial distances used for the numerical and experimental evaluations of the diffractive decoders reported in this work.

The diffractive layers were constructed using diffractive features, each with dimensions of $0.667\lambda$, arranged in a $66.7\lambda \times 66.7\lambda$ grid ($50\ mm \times 50\ mm$). Similarly, we established an effective pixel size of $1.33\lambda$ at the measurement plane, while both the phase-only wavefront modulator and the output field of view were designed as $40\lambda \times 40\lambda$ ($30\ mm \times 30\ mm$). The LR wavefront modulator's pixel pitch was set to $8\lambda \times 8\lambda$, yielding a PSR factor of 6.

The diffractive components were fabricated using a 3D-printing material characterized by a complex refractive index of $\tau(\lambda) \approx 1.6518 + j0.0612$ at the operational wavelength, $\lambda \approx 0.75\ mm$. The thickness of each diffractive feature, denoted as $h_l[m,n]$, was optimized within the range of $[0.5\ mm, 1.644\ mm]$, corresponding to a phase modulation span of $[-\pi, \pi)$.

To enhance the robustness of our system, we implemented a vaccination process[33] during the training. This involved introducing random alignment errors to the positions of diffractive layers and the input patterns during the training phase. The random errors are represented by $D^l = (D_x, D_y, D_z)$, indicating deviations of the diffractive layer $l$ from its ideal position. Here, $D_x$, $D_y$, and $D_z$ independently follow a uniform distribution, such that $D_x \sim U(-\Delta_x, \Delta_x)$, $D_y \sim U(-\Delta_y, \Delta_y)$, and $D_z \sim U(-\Delta_z, \Delta_z)$, where the parameters $\Delta_x$, $\Delta_y$, and $\Delta_z$ denote the hyperparameters of the possible alignment errors along the x-, y-, and z-axes, respectively. Consequently, the position of the diffractive layer $l$ at the $i^{th}$ iteration, denoted as $L^{(l,i)}$, is expressed as:

$$L^{(l,i)} = \left(L_x^l, L_y^l, L_z^l\right) + \left(D_x^{(l,i)}, D_y^{(l,i)}, D_z^{(l,i)}\right) \qquad (8)$$

The alignment error ranges for our experiments were set to $0.67\lambda$ for both $\Delta x$ and $\Delta y$, and $0.534\lambda$ for $\Delta z$.

Following the decoder optimization, we converted the thickness maps of the diffractive layers and the corresponding phase-encoded representations into STL files using MATLAB. These designs were then physically fabricated using an Objet30 Pro 3D printer (Stratasys). To precisely adjust the distance between the decoder and the fixed terahertz wave detector at the output plane, a slidable structure, illustrated in **Fig. 5b**. was designed and 3D-printed. This configuration enables output image measurements over an extended DOF by controlling the distance from the diffractive layer to the output plane ($z_{3,i}$).

**Experimental setup**

The experimental apparatus is depicted in **Fig. 5c**. At its core, a Virginia Diodes Inc. WR2.2 modular amplifier/multiplier chain (AMC) serves as the radiation source. This device transforms an 11.111 GHz (fRF1) input, amplified to 10 dBm, into a 0.4 THz continuous wave through a 36-fold frequency multiplication process. To enhance detection sensitivity, the output undergoes 1-kHz amplitude modulation. Terahertz waves emanate from a diagonal horn antenna, traversing 60 cm before encountering the target object and, subsequently, a 3D-printed diffractive decoder. A Virginia Diodes Inc. single-pixel Mixer/AMC, operating with an 11.083 GHz (fRF2) local oscillator at 10 dBm, captures and down-converts the diffracted field to a 1 GHz intermediate frequency. A detector ($0.5 \text{ mm} \times 0.25 \text{ mm}$) is mounted on Thorlabs NRT100 motorized linear stages, enabling precise X-Y positioning. It scanned the field of view in 0.75 mm steps, and subsequently, $3 \times 3$ bilinear upsampling and 4×4-pixel binning were employed to increase the signal-to-noise ratio and approximate the system's ~$1.33\lambda$ (1 mm) output pixel size. Mini-Circuits ZRL-1150-LN+ amplifiers provide 40 dB gain, followed by a KL Electronics 3C40-1000/T10-O/O bandpass filter (1 GHz ± 10 MHz). An HP 8495B attenuator allows for system calibration before the signal reaches a Mini-Circuits ZX47-60 power detector. Final processing occurs in a Stanford Research SR830 lock-in amplifier synchronized to the 1 kHz modulation. The resulting data undergoes linear scaling and dynamic range adjustment, where the extreme 5% of values at both ends are saturated, with the remainder mapped to a 0-1 range.

**Supplementary Information** includes:

- **Supplementary Figures 1-2**
- **Loss function and DOF-aware optimization strategy**

# Figures

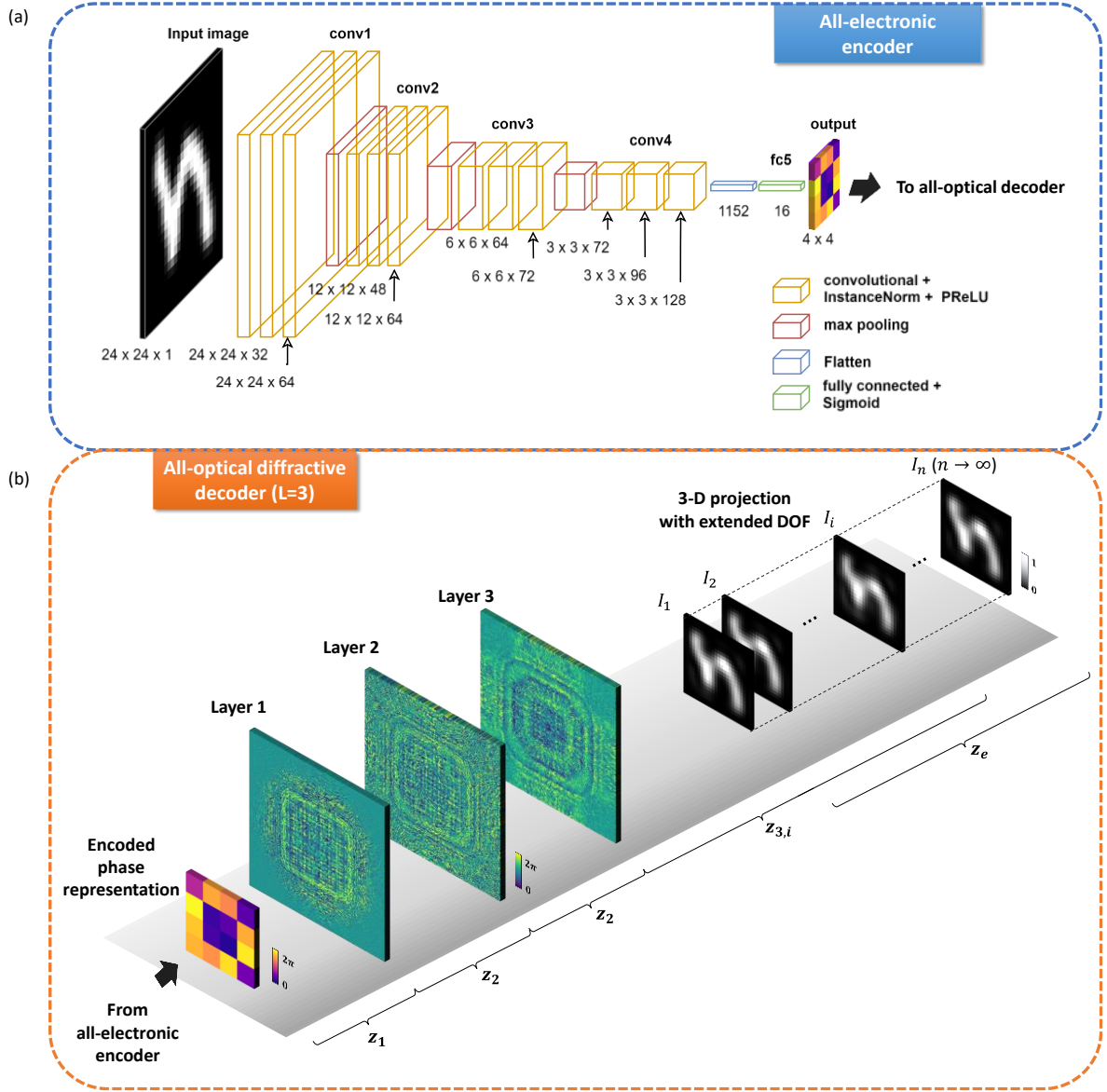

**Figure 1 Schematic illustration of the hybrid PSR image projection system with a digital encoder and a diffractive optical decoder.** (a) The digital encoder transforms grayscale input images (each with $24 \times 24$ pixels) into compact phase-encoded representations of size $\frac{24}{k} \times \frac{24}{k}$ pixels, where $k = 6$ is shown here as an example. (b) The diffractive decoder all-optically decodes and modulates the wavefront to achieve a continuous projection with an extended depth of field, EDOF ($z_e$). The distance $z_1$ represents the axial distance from the input plane to the first diffractive layer; $z_2$ denotes the axial spacing between adjacent passive diffractive layers. $z_{3,1}$ and $z_{3,n}$ indicate the axial distances from the last diffractive layer to the nearest and farthest projection planes, respectively. For demonstration, discrete positions within this *continuous* projection field are shown, denoted as $z_{3,1}, z_{3,2}, \ldots, z_{3,i}, \ldots, z_{3,n}$, with the corresponding cross-sectional images $I_1, I_2, \ldots, I_i, \ldots, I_n$, respectively.

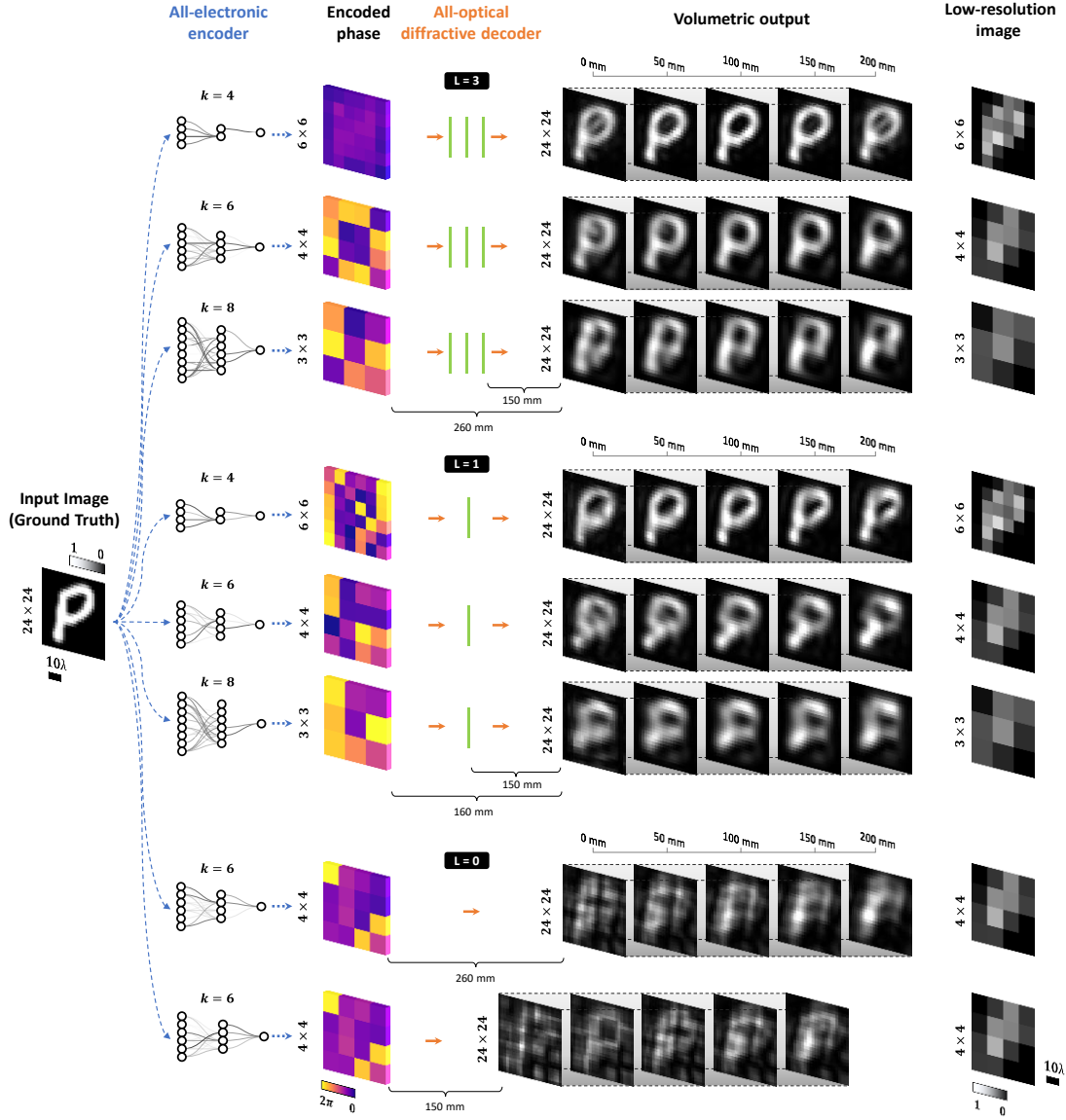

**Figure 2 Performance comparison of hybrid PSR image projection systems.** The input images (each 24 × 24 pixels) are processed through our PSR image projection systems trained for varying PSR factors ($k = 4, 6, 8$) and different numbers of passive diffractive layers ($L = 1, 3$) for the diffractive decoder. Each row represents a distinct system configuration with jointly trained encoder-decoder pairs, optimized independently from other configurations. The volumetric outputs are continuous projections with an extended DOF, i.e., $z_e = 200$ mm (~266.85λ). For illustration purposes, several propagation distances, i.e., 0 mm, 50 mm (~66.71λ), 100 mm (~133.43λ), 150 mm (~200.14λ), 200 mm (~266.85λ), are selected to showcase cross-sectional views. The rightmost column displays the low-resolution images corresponding to each PSR factor, $k$. The bottom two rows illustrate system configurations with *no* diffractive layers (the encoder-only cases, where $L = 0$), serving as a comparison, highlighting the significance of the diffractive decoder.

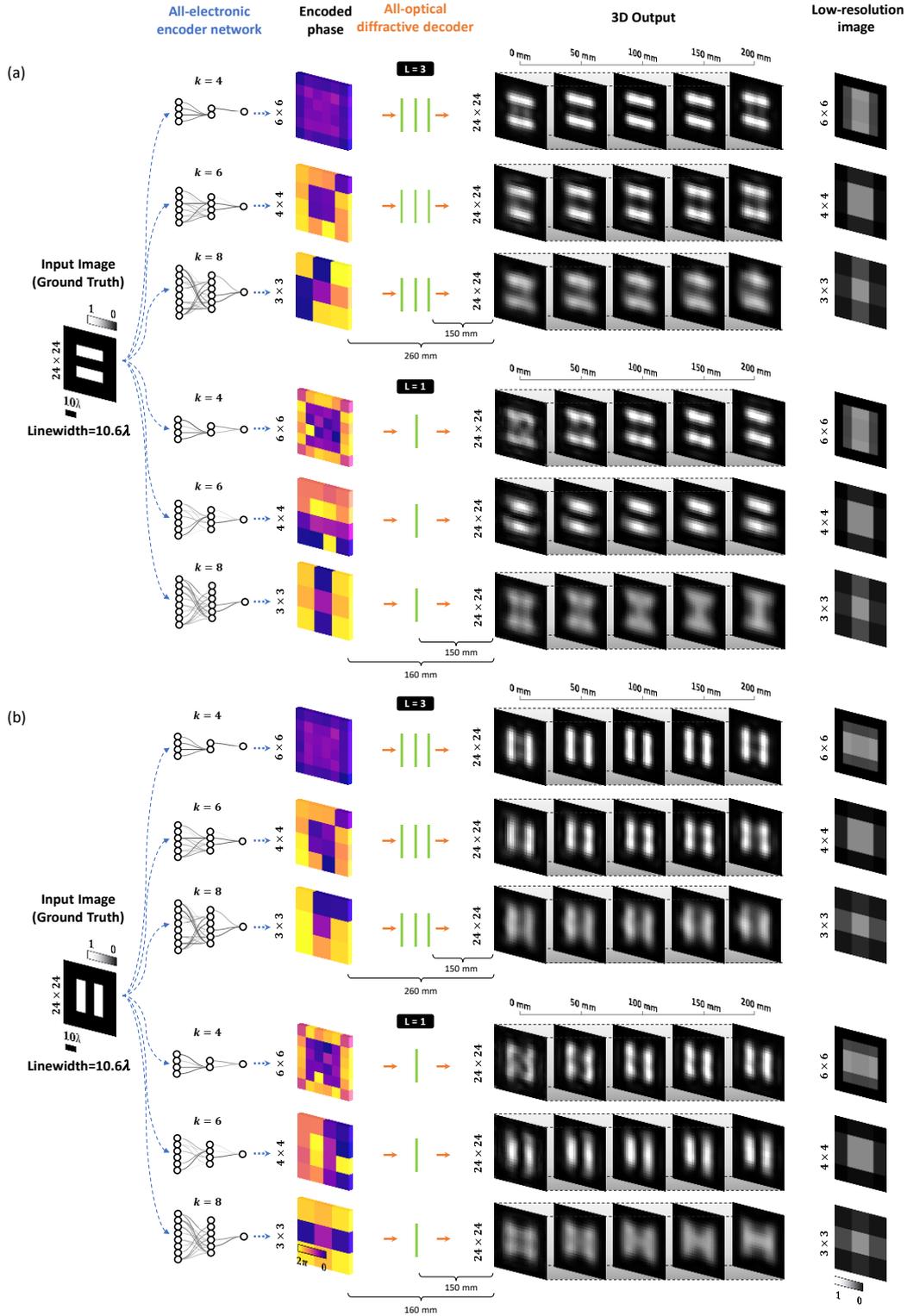

**Figure 3 External generalization performance of hybrid PSR image projection systems on grating patterns – never seen before.** This figure demonstrates the efficacy of the system in generalizing to unseen data types, utilizing encoder-decoder models trained with the EMNIST dataset for each combination of $k$ and $L$. (a) Horizontal and (b) vertical grating inputs (24 × 24 pixels) are processed through each configuration. Encoded phase patterns and volumetric outputs, axially sampled at 0 mm, 50 mm (~66.71λ), 100 mm (~133.43λ), 150 mm (~200.14λ), 200 mm (~266.85λ)

showcase continuous projection over a DOF of 200 mm (~266.85λ). The rightmost column presents the corresponding low-resolution images for each PSR factor, $k$.

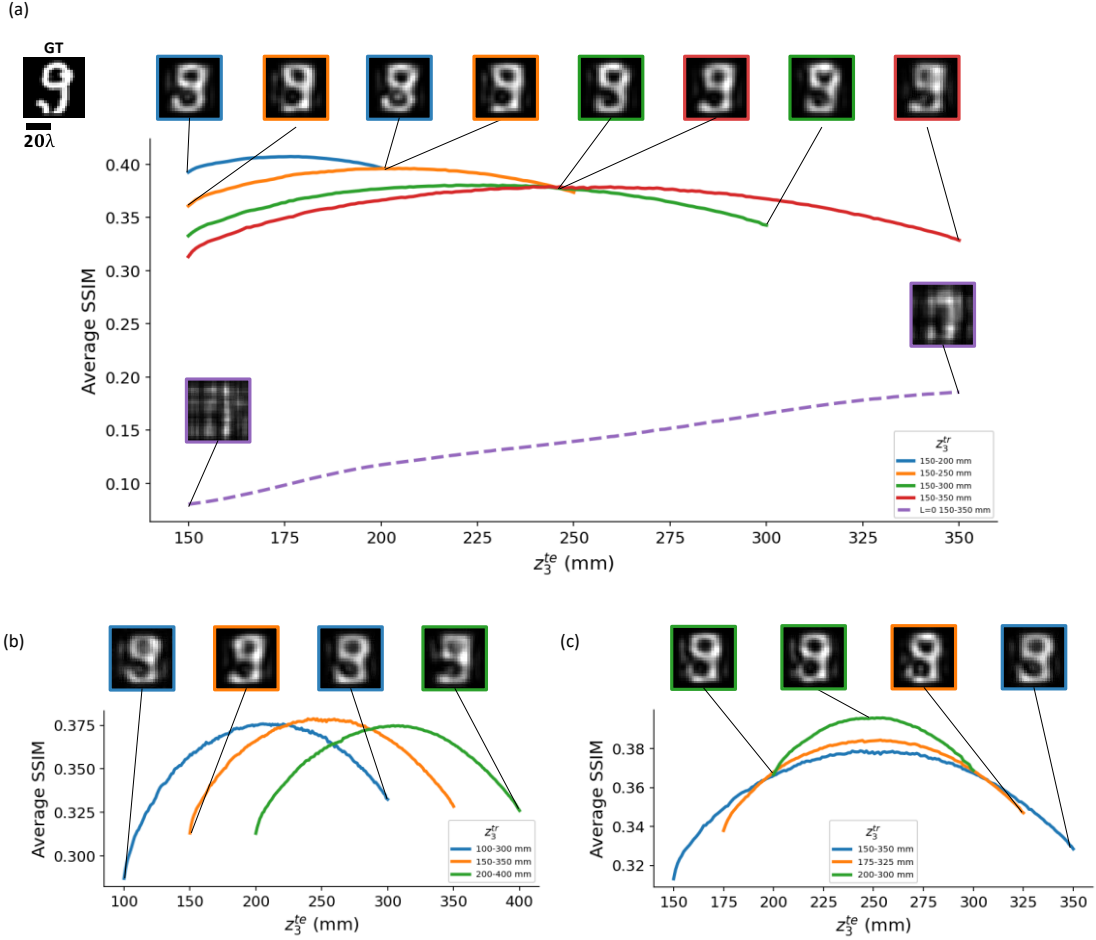

**Figure 4 Quantitative analysis of the hybrid PSR image projection systems across varying $z_3$ ranges.** Each solid curve represents a hybrid system ($L = 3$, $k = 6$) trained over a specific $z_3^{tr}$ training range and evaluated over the corresponding $z_3^{te}$ testing range, as indicated in the legend. The dashed purple curve depicts the performance of a projection system without a diffractive decoder (i.e., $L = 0$), trained and evaluated over an equivalent axial depth range. Representative reconstructed images at selected $z_3^{te}$ positions are displayed above the corresponding curves. (a) Impact of increasing the depth range on the image projection fidelity, with all ranges ($z_3^{te}$) starting from 150 mm (~200.14λ). (b) Performance comparison for depth ranges of a constant width $z_e$ of 200 mm (~266.85λ) but varying central positions, elucidating the influence of range placement on the PSR image projection quality. (c) Average SSIM as a function of $z_3^{te}$ for axial ranges sharing a common midpoint at 250 mm (~333.56λ) but differing in width, $z_e$.

(a) Phase profile of the 1-layer diffractive decoder

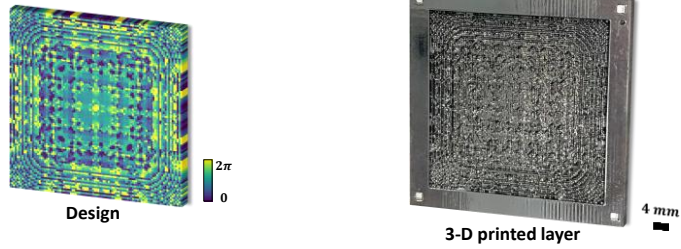

(b) Slidable structure for EDOF display

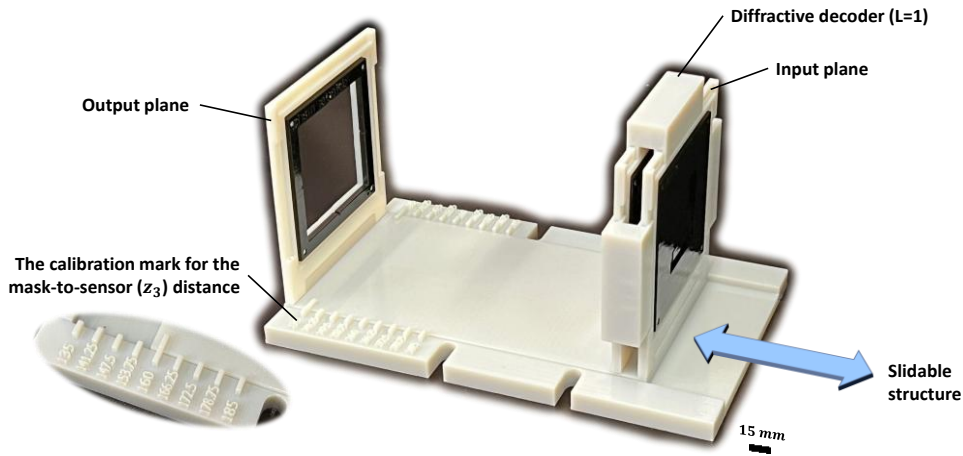

(c) Experimental setup with the projection system

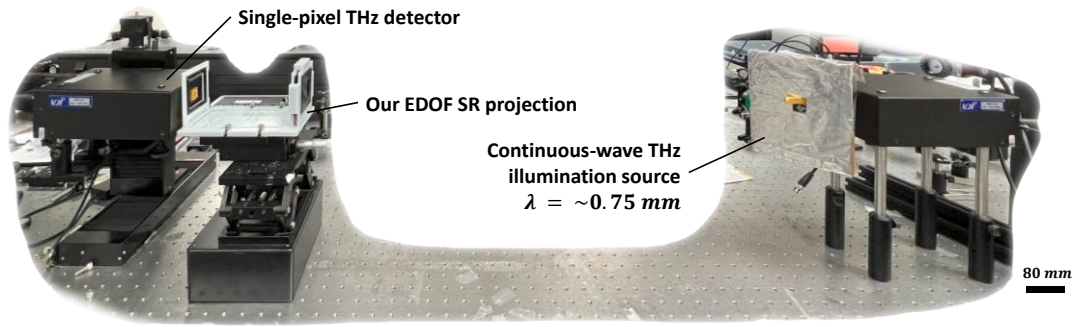

**Figure 5 Experimental setup for EDOF PSR image projection system with $L = 1$.**
(a) The phase profile of the designed diffractive decoder and its 3D fabricated version. (b) This 3D printed structure (white color) facilitates the relative axial translation of the diffractive decoder with respect to the terahertz detector at the output plane, enabling a variable layer-to-sensor distance to measure the resulting image projections at different axial positions. A calibration mark is integrated into the structure to ensure accurate measurement of the $z_3$ distance. (c) Experimental setup employing a continuous wave terahertz illumination and a THz detector.

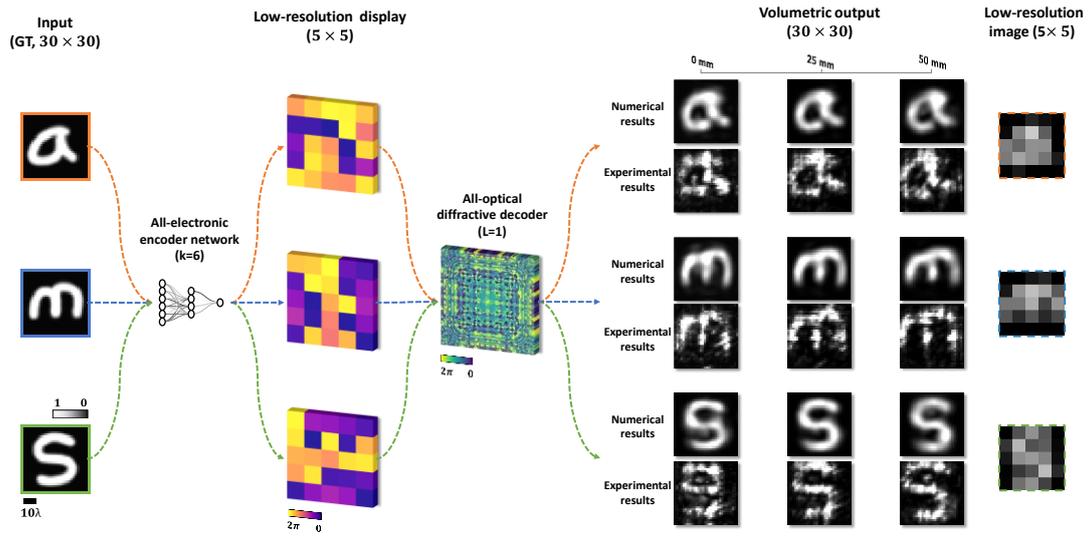

**Figure 6 Experimental results of the extended DOF PSR image projection system.** The hybrid system combines an all-electronic encoder network ($k = 6$) with an all-optical diffractive decoder ($L = 1$). Input images ($30 \times 30$ pixels) are first encoded into low-resolution ($5 \times 5$ pixels) phase patterns, which are then processed by the diffractive decoder through wave propagation. The volumetric output ($30 \times 30$ pixels) shows both the numerical and the experimental images at selected depths of 0 mm, 25 mm (~33.36λ), and 50 mm (~66.71λ), over a continuous volumetric projection. The rightmost column shows the corresponding low-resolution ($5 \times 5$ pixels) images for comparison.